\documentclass[12pt,reqno]{amsart} 
\usepackage{amssymb}\usepackage{epsf}

\numberwithin{equation}{section}

\theoremstyle{definition}

\def\stackreb#1#2{\ \mathrel{\mathop{#1}\limits_{#2}}}

\newcommand{\beq}{\begin{equation}}
\newcommand{\be}{\begin{equation}}
\newcommand{\ee}{\end{equation}}
\newcommand{\ba}{\begin{eqnarray}}
\newcommand{\ea}{\end{eqnarray}}

\newcommand{\lab}[1]{\label{#1}}

\newcommand{\C}{\mathbb C}

\newcommand{\Z}{\mathbb Z}

\newcommand{\PP}{\mathbb P}
\newcommand{\V}{\mathbb V}
\newcommand{\T}{\mathbb T}

\begin{document}

\title
{The rarefied elliptic Bailey lemma and the Yang-Baxter equation}

\author{V. P. Spiridonov}

\address{Laboratory of theoretical physics, Joint Institute for Nuclear
Research, Dubna, Moscow reg., 141980, Russia and
St. Petersburg Department of the Steklov Mathematical Institute
of Russian Academy of Sciences,
Fontanka 27, St. Petersburg, 191023 Russia}
\begin{abstract}
An elliptic Bailey lemma is formulated on the basis of the
univariate rarefied elliptic beta integral. It leads to a generalized operator star-triangle
relation and a new solution of the Yang-Baxter equation written as an integral operator
with a rarefied elliptic hypergeometric kernel.
\end{abstract}

\maketitle
%

\section{Introduction}

The Bailey lemma techniques is a well known tool for constructing various nontrivial relations
for $q$-hypergeometric series \cite{aar}. Originally it was developed for proving Rogers-Ramanujan
type identities which later appeared in statistical mechanics of two-dimensional
lattice systems \cite{bax:book}. However, it has found a number of other applications in mathematics
and mathematical physics \cite{Warnaar}. In \cite{spi:bailey2} this techniques was generalized
to integrals on the basis of the elliptic beta integral evaluation formula established in \cite{spi:umn}.
It is based on the elliptic analogue of the Fourier transformation whose inversion
properties were described in \cite{SW}.
As shown in \cite{DS} the key identity for operator elements of the corresponding
elliptic Bailey lemma has the meaning of the star-triangle relation. This allows
construction of the most complicated known integral operator solution of the
Yang-Baxter equation and the use of full machinery of the quantum inverse scattering
method \cite{TF} for investigating the physical system behind it.

Nowadays elliptic hypergeometric functions form a well established set of special
functions with many beautiful properties \cite{spi:aar}. Nevertheless their theory,
still being sufficiently young, continues to develop in various, sometimes unexpected
directions. In particular, coincidence of elliptic hypergeometric integrals
with superconformal indices of supersymmetric four-dimensional field theories \cite{RR}
brought a qualitatively new understanding of the structure of these functions.
One of the recent new directions was inspired by the consideration of
supersymmetric field theories  on the space-time involving a particular lens space.
The first rigorous result in this direction was obtained by Kels \cite{kels}
who generalized the elliptic beta
integral of \cite{spi:umn} to a computable finite sum of integrals containing additional
integer variables. In \cite{spi:rare} the author has proved this formula by a slightly
different method and established its extension to half-integer values of the discrete variables.
An equivalent result was obtained also in \cite{KY} as a subcase of a symmetry transformation
for multivariate functions. The most complicated known exact evaluation formula for
such ``rarefied" elliptic hypergeometric functions was described in \cite{spi:rare},
which can be considered as a generalization of the Selberg integral \cite{aar}.

An application of the rarefied elliptic beta integral to two-dimensional statistical
mechanics models was considered in \cite{kels}. It extends the considerations of \cite{BMS,BS,stat}
of two-dimensional spin lattice systems associated with the standard elliptic beta integral
and its degenerations. In this paper we generalize the univariate Bailey lemma of \cite{spi:bailey2}
to the case of rarefied elliptic beta integral evaluation formula and build the
corresponding solution of the Yang-Baxter equation following the considerations of \cite{SD,DM,DS}.

We start our considerations from presenting notations for the standard elliptic
hypergeometric functions. For $|p|<1$ we define a Jacobi theta function as
$$
\theta(z;p):=(z;p)_\infty(pz^{-1};p)_\infty,\quad
(z;p)_\infty:=\prod_{j=0}^\infty(1-zp^j), \quad  z\in\ \mathbb{C}^\times.
$$
The standard odd Jacobi theta function is related to it as follows
\begin{eqnarray} \lab{theta1} &&
 \theta_1(u|\tau)=-\theta_{11}(u)=-\sum_{k\in \mathbb Z}e^{\pi {i} \tau(k+1/2)^2}
e^{2\pi {i} (k+1/2)(u+1/2)}
\\  && \makebox[4em]{}
={i}p^{1/8}e^{-\pi {i} u}(p;p)_\infty\theta(e^{2\pi {i} u};p),
\quad p=e^{2\pi {i} \tau}.
\nonumber\end{eqnarray}
One has the symmetry properties
\beq
\theta(pz;p)=\theta(z^{-1};p)=-z^{-1}\theta(z;p)
\lab{thetasym}\ee
and the Laurent series expansion
\beq
\theta(z;p)=\frac{1}{(p;p)_\infty} \sum_{k\in\Z} (-1)^kp^{k(k-1)/2}z^k.
\lab{Jtri}\ee

The standard elliptic gamma function has the form
\begin{equation}
\Gamma(z;p,q):=
\prod_{j,k=0}^\infty\frac{1-z^{-1}p^{j+1}q^{k+1}}{1-zp^{j}q^{k}},
\qquad |p|, |q|<1,\quad z\in\mathbb{C}^\times,
\label{eg}\end{equation}
and satisfies the equations
$$
\Gamma(z;p,q)=\Gamma(z;q,p),\qquad \Gamma(pq/z;p,q)\Gamma(z;p,q)=1
$$
and relations
$$
\Gamma(qz;p,q)=\theta(z;p)\Gamma(z;p,q), \quad
\Gamma(pz;p,q)=\theta(z;q)\Gamma(z;p,q).
$$

We shall need also the elliptic gamma function of the second order
$$
\Gamma(z;p,q,t)=\prod_{j,k,l=0}^\infty (1-zp^jq^kt^l)
(1-z^{-1} p^{j+1}q^{k+1}t^{l+1}), \quad |t|,|p|,|q|<1,\; z\in\mathbb{C}^\times,
$$
which is symmetric in all three bases $p,q,t$. It satisfies the relations
\beq
\Gamma(qz;p,q,t)=\Gamma(z;p,t)\Gamma(z;p,q,t), \qquad \Gamma(pqtz;p,q,t)=\Gamma(z^{-1};p,q,t).
\label{2ndordereqn}\ee

\section{Rarefied elliptic beta integral}

A generalization of the elliptic gamma function used in \cite{kels}, which was called the
rarefied elliptic gamma function in \cite{spi:rare}, has the following form
\begin{eqnarray}\label{reg} &&
\gamma^{(r)}(z,m;p,q):=\Gamma(zp^m;p^r, pq)\Gamma(zq^{r-m};q^r, pq),
\end{eqnarray}
where one has two integer parameters $r\in\mathbb{Z}_{>0}$, $m\in\mathbb{Z}$, and
$\Gamma(z;p,q)$ is the standard elliptic gamma function \eqref{eg}.
The function \eqref{reg} can be written as a ``rarefied" product \cite{spi:rare}
\beq
\gamma^{(r)}(z,m;p,q)
=\prod_{k=0}^{m-1}\Gamma(q^{r-m}z(pq)^k;p^r,q^r)\prod_{k=0}^{r-m-1}\Gamma(p^mz(pq)^k;p^r,q^r),
\label{regm_r1}\ee
valid for $0\leq m\leq r$. For other values of $m$ similar representation is obtained
from the quasiperiodicity property
\beq
\gamma^{(r)}(z,m+r;p,q) =(-z)^{-m} q^{ m(m+1)/2}p^{-m(m-1)/2}\gamma^{(r)}(z,m;p,q).
\label{r-per}\ee
One has also the permutational symmetry
\beq
\gamma^{(r)}(z,m;p,q)=\gamma^{(r)}(z,r-m;q,p)
\ee
and the inversion relation
\beq
\gamma^{(r)}(z,m;p,q)\gamma^{(r)}(\textstyle{\frac{pq}{z}},r-m;p,q)=1.
\label{inv0}\ee

In \cite{spi:rare} the following normalized function was used for proving
various exact relations for rarefied elliptic hypergeometric functions
\beq
\Gamma^{(r)}(z,m;p,q):=
(-z)^{\frac{m(m-1)}{2}}p^{\frac{m(m-1)(m-2)}{6}}q^{-\frac{m(m-1)(m+1)}{6}}\gamma^{(r)}(z,m;p,q).
\label{regm}\ee
For $r=1$, one has $\Gamma^{(1)}(z,m;p,q)=\Gamma(z;p,q)$ independently of $m$.
This function has the permutational symmetry $\Gamma^{(r)}(z,m;p,q)=\Gamma^{(r)}(z,-m;q,p)$
and the inversion relation $\Gamma^{(r)}(z,m;p,q)\Gamma^{(r)}(\textstyle{\frac{pq}{z}},-m;p,q)=1$.
However, it remains a quasiperiodic function similar to \eqref{r-per}.

Following the suggestion of \cite{spi:rare} to use analytical dependence
on the discrete variables, in \cite{GK} the following normalized rarefied elliptic
gamma function was introduced
\beq
\Gamma(u,m;\tau,\sigma) :=
e^{\pi i \frac{m(m-r)}{2r}(2u-\tau-\sigma+\frac{1}{3}(2m-r)(\tau-\sigma-1))}
\gamma^{(r)}(z,m;p,q),
\label{REGper}\ee
where $z=e^{2\pi i u},\, p=e^{2\pi i \tau}, \, q=e^{2\pi i\sigma}$.
The key property of this normalization choice is the periodicity
\beq
\Gamma(u,m+r;\tau,\sigma) = \Gamma(u,m;\tau,\sigma).
\label{period}\ee
However, it is not a single valued function of $z, p,$ and $q$ and
the permutational symmetry is modified to
\beq
\Gamma(u,-m;\sigma,\tau)= e^{\pi i \frac{m(m-r)(2m-r)}{3r}}\Gamma(u,m;\tau,\sigma).
\label{symmetry}\ee
The inversion relation has a compact form
\beq
\Gamma(u,m;\tau,\sigma)\Gamma(\tau+\sigma-u,-m;\tau,\sigma)=1.
\label{inv}\ee
The recurrence relations take the form
\begin{eqnarray} \label{recurrence1} && \makebox[-3em]{}
\frac{\Gamma(u+\sigma,m+1;\tau,\sigma)}{\Gamma(u,m;\tau,\sigma)}
=e^{\frac{\pi i}{2r}((2u-1)(2m+1-r)+(\sigma-\tau+1)\frac{1-r^2}{3}+2(\tau-1)m(m-r))}\theta(zp^{m};p^r),
\\ && \makebox[-3em]{}
\frac{\Gamma(u+\tau,m-1;\tau,\sigma)}{\Gamma(u,m;\tau,\sigma)}
= e^{\frac{\pi i}{2r}((2u+1)(-2m+1+r)-(\sigma-\tau+1)\frac{1-r^2}{3}+2(\sigma+1)m(m-r))}\theta(zq^{-m};q^r).
\label{recurrence2} \end{eqnarray}
The limiting relation
\beq\label{rfres} \qquad
\stackreb{\lim}{u\to 0}(1-e^{2\pi i u}) \Gamma(u,0;\tau,\sigma)
=\stackreb{\lim}{z\to 1}(1-z) \gamma^{(r)}(z,0;p,q)
=\frac{1}{(p^r;p^r)_\infty(q^r;q^r)_\infty}
\ee
is used for the residue calculus.

The rarefied version of the elliptic beta integral \cite{spi:umn} has the following form.
Let complex parameters $t_1, \dots ,t_6,p,q $ and discrete variables
$$
n_1,\ldots, n_6\in\mathbb{Z}+\mu, \quad \mu = 0 \ \text{or}\ \frac{1}{2},
$$
satisfy  the constraints $|t_a|,|p|,|q| <1$ and the balancing condition
\beq
\prod_{a=1}^6 t_a=pq, \qquad \sum_{a=1}^6 n_a=0.
\label{balance} \ee
Then
\begin{equation} \label{rfint}
\kappa^{(r)}\sum_{m\in \mathbb{Z}_r+\mu}
\int_{\mathbb{T}}\rho^{(r)}(z,m;t_a,n_a)
\frac{dz}{2\pi {i}z}
= \prod_{1 \leq a < b \leq 6} \Gamma^{(r)}(t_a t_b,n_a+n_b;p,q),
\end{equation}
where $\mathbb T$ is the positively oriented unit circle and $\mathbb{Z}_r=0,1,\ldots, r-1$,
$$
\kappa^{(r)}=\frac{1}{2}(p^r;p^r)_\infty (q^r;q^r)_\infty,
$$
and the integrand has the form
\beq
\rho^{(r)}(z,m;t_a,n_a):=
\frac{\prod_{a=1}^6\Gamma^{(r)}(t_az^{\pm 1},n_a\pm m;p,q) }
{\Gamma^{(r)}(z^{\pm2},\pm 2m;p,q)}.
\label{ker_rebeta}\ee
Here we use the compact notation
\beq
\Gamma^{(r)}(tz^{\pm1},n\pm m;p,q):=\Gamma^{(r)}(tz,n+m;p,q)
\Gamma^{(r)}(tz^{-1},n-m;p,q).
\ee

For $\mu=0$ this evaluation was established by Kels in \cite{kels} and for $\mu =1/2$ in
\cite{spi:rare} and \cite{KY} (in a different form as $A_1\to A_0$ symmetry transformation).

In terms of the periodic elliptic gamma function
\begin{equation} \label{rfintadd}
\kappa^{(r)}\sum_{m\in \mathbb{Z}_r+\mu}
\int_0^1\rho(u,m;s_a,n_a)du
= \prod_{1 \leq a < b \leq 6} \Gamma(s_a+s_b,n_a+n_b;\tau,\sigma),
\end{equation}
where
\beq
\rho(u,m;s_a,n_a):=
\frac{\prod_{a=1}^6\Gamma(s_a\pm u,n_a\pm m;\tau,\sigma) }
{\Gamma(\pm 2 u,\pm 2m;\tau,\sigma)}
\label{ker}\ee
with  $\text{Im}(s_a)>0$, $\sum_{a=1}^6 s_a=\tau+\sigma, \, \sum_{a=1}^6 n_a=0 \mod r,$
and the compact notation
\beq
\Gamma(s\pm u,n\pm m;\tau,\sigma):=\Gamma(s+u,n+m;\tau,\sigma)
\Gamma(s- u,n-m;\tau,\sigma).
\ee
It appears that both normalization factors for the rarefied elliptic gamma function
yield equivalent generalizations of the elliptic beta integral evaluation formula.

Let us denote the left-hand side of equalities in \eqref{rfint} or \eqref{rfintadd}
as the sum $\sum_{m=0}^{r-1}c_m$. Then, as shown in \cite{spi:rare}, this sum
can be reduced for $\mu=0$ to
\beq
\sum_{m=0}^{r-1}c_m =
\left\{
\begin{array}{cl}
c_0+c_{r/2}+2\sum_{m=1}^{r/2-1}c_m & \text{for even}\, r, \\
c_0+2\sum_{m=1}^{(r-1)/2}c_m  & \text{for odd}\, r,
\end{array}
\right.
\label{e=0red}\ee
and for $\mu=1/2$ to
\beq
\sum_{m=0}^{r-1}c_m =
\left\{
\begin{array}{cl}
2\sum_{m=0}^{r/2-1}c_m & \text{for even}\, r, \\
c_{(r-1)/2}+2\sum_{m=0}^{(r-3)/2}c_m  & \text{for odd}\, r.
\end{array}
\right.
\label{e=1red}\ee

\section{The rarefied elliptic Bailey lemma}

In \cite{spi:bailey2} an elliptic analogue of the Fourier transformation was introduced in the context of
an integral generalization of the Bailey chains technique \cite{Warnaar}.
We would like to introduce a rarefied analogue of this transformation with the
help of the periodic rarefied/lens elliptic gamma function \eqref{REGper}.
Let us take two complex variables $v,u\in\C$, and three discrete variables
$m\in\Z_r+\mu,$ $k\in\Z_r+\nu,$ $\mu, \nu=0,\frac{1}{2}$, and $n\in\Z+\mu+\nu$
($\nu$ is a variable independent of $\mu$).
It is convenient to introduce parities $p(m)=(-1)^{2\mu},\, p(k)=(-1)^{2\nu}$,
so that $p(n)=p(m)p(k)$. Now we define the following integral transformation
\begin{eqnarray}\nonumber
&& \beta(v,k;s,n)=M(s,n)_{v,k;u,m}\alpha(u,m;s,n)
\\ && \makebox[2em]{}
=\kappa^{(r)}\int_{-1/2}^{1/2}\sum_{m\in\mathbb{Z}_r+\mu}
\frac{\Gamma(s\pm v\pm u, n\pm k \pm m)}
{\Gamma(2s,2n)\Gamma(\pm2 u,\pm 2m)}\alpha(u,m;s,n)du
\label{REFT}\end{eqnarray}
with the assumption that $\text{Im}(s \pm v)>0$. For convenience we impose also the constraints
$|\text{Re}(s\pm v)|<\frac{1}{2}$ which fix positions of poles in the integrand of \eqref{REFT}.
Here and below we use the convention
$$
\Gamma(s\pm v\pm u, n\pm k \pm m):=\prod_{\epsilon,\delta=\pm1}
\Gamma(s+\epsilon v+\delta u, n+\epsilon k +\delta m).
$$
Dependence on the modular parameters $\tau, \sigma$ and the integer variable $r$ is suppressed
for brevity. Because of the periodicity property \eqref{period} one can naturally
assume that $0\leq m<r$ (as in \eqref{REFT}) and $0\leq k<r$.

The constraints on continuous parameters can be relaxed by analytic continuation,
which is reached by deforming the contour of integration without crossing integrand
singularities. The pairs of functions connected by \eqref{REFT} will be called
Bailey pairs with respect to the parameters $s$ and $n$. Using the evaluation formula
\eqref{rfintadd} one can find a particular explicit Bailey pair.
This is the standard terminology in the theory of Bailey chains \cite{aar,Warnaar}.
The definition \eqref{REFT} leads to a natural generalization
of the integral Bailey chains technique which was suggested in \cite{spi:bailey2}.

Note that in the operator $M(t,n)_{v,k;u,m}$ one can have $n=0$ only if $\mu=\nu$,
or $p(k)=p(m)$. In this case one has the following limiting relation for sequences of functions
$f(u,m)$ holomorphic near the cut $u\in [-\frac{1}{2},\frac{1}{2}]$:
\begin{equation}
\lim_{s\to 0} M(s,0)_{v,k;u,m}f(u,m)=\frac{1}{2}(f(v,k)+f(-v,\tilde k)),
\label{unit}\end{equation}
where $\tilde k=r-k$ for $k>0$ and $\tilde k=0$ for $k=0$.

In order to prove this statement let us consider in detail the divisor of the
integrand in \eqref{unit}. It is convenient to do in terms of the variables
$z=e^{2\pi i u},\, w=e^{2\pi i v},\, t=e^{2\pi i s}$ and bases $p, q$ since the non-analytical parts
in \eqref{ker} have only branching singularities.
At first glance, sequences of poles of the integrand converging to $z=0$ point are located at
$$
P_{\rm in}^A= \{tw^{\pm1}q^ip^{\pm k-m+i+rj}\},\quad
 P_{\rm in}^B= \{tw^{\pm1}q^{r\mp k+m+i+rj}p^i\}
$$
with $i,j \in\mathbb{Z}_{\geq 0}$, and those going to infinity seat at the points
$$
P_{\rm out}^A = \{t^{-1}w^{\mp1}q^{-i}p^{\mp k-m-i-rj}\},\quad
P_{\rm out}^B = \{t^{-1}w^{\mp1}q^{\pm k+m-i-r(j+1)}p^{-i}\}.
$$
Note that $P_{\rm in}$ and $P_{\rm out}$ are not identically reciprocal to each other.
Analogously, possible zeros of the integrand converge to $z=0$ by the points
$$
Z_{\rm in}^A = \{t^{-1}w^{\mp1}q^{i+1}p^{\mp k-m+i+1+r(j+1)}\},\quad
Z_{\rm in}^B = \{t^{-1}w^{\mp1}q^{\pm k+m+i+1+rj)}p^{i+1}\}
$$
where  $i,j \in\mathbb{Z}_{\geq 0}$, and go to infinity along the points
$$
Z_{\rm out}^A= \{tw^{\pm1}q^{-i-1}p^{\pm k-m-i-1-r(j+1)}\},\quad
Z_{\rm out}^B= \{tw^{\pm1}q^{\mp k+m-i-1-rj}p^{-i-1}\}.
$$
The sets $Z_{\rm in}$ and $Z_{\rm out}$ are also not exactly reciprocal to each other.

Looking at the structure of these sets one can see that potentially
there may be nontrivial cancellations of poles and zeros.
Remind that without loss of generality we have limited values of the discrete
variable $k$ to $0\leq k <r$. Consider possible overlap
of $P_{\rm in}^B$ with $Z_{\rm out}^A$. Equating positions of corresponding poles and zeros
we see that this may happen if $rj+i+i'=-r\pm k-m-1$
with $i, i', j\in \Z_{\geq 0}$. However, for the taken values of $k$ this equation
has no solutions and the maximally close to $\T$ pole from $P_{\rm in}^B$ is either $twq$
or $tw^{-1}q^{r+2\mu}$.

Consider now the possible overlap of $P_{\rm in}^A$ with $Z_{\rm out}^B$. The poles
$\{twq^ip^{k-m+i+rj}\}$ may be cancelled if $rj+i+i'=m-k-1$. Let $m\leq k$, then this
equation has no solutions and the closest to $\T$ pole is $z=tw$, which is reached for $m=k$.
For $k<m$ there are nontrivial solutions $j=0,\, i,i'=0,\ldots,m-k-1$, showing that
all potential poles lying outside of circle of radius $|tw|$ are cancelled by the zeros
from $Z_{\rm out}^B$. In the consideration of poles $\{tw^{-1}q^ip^{-k-m+i+rj}\}$
one can see that the closest to $\T$ pole $z=tw^{-1}$ is reached for $m=k=0$
or $m=r-k$ for $k>0$ (in this case one can replace $-k$ by $r-k$ from $r$-periodicity in $k$).
All other potential poles
lying outside of the circle of radius $|tw^{-1}|$ cancel with zeros.

The poles associated with the point $t^{-1}w^{-1}$ from the set $P_{\rm out}^A$ cannot
be cancelled by zeros from $Z_{\rm in}^B$, since the equation $rj+i+i'=-k-m-1$
has no solutions. As a result, the closest to $\T$ pole $t^{-1}w^{-1}$ can be reached
only for $m=k=0$. As to the $t^{-1}w$-point related poles, they may cancel, if $rj+i+i'=k-m-1$.
For $m\geq k$ this is not possible and the closest to $\T$ pole $z=t^{-1}w$ is reached for $m=k$.
For $m<k$ a nontrivial cancellation of poles and zeros takes place and there do not appear new closest
to $\T$ poles. Finally, the $t^{-1}w^{-1}$-point related poles from $P_{\rm out}^B$ can
overlap with $Z_{\rm in}^A$, if $rj+i+i'=k+m-r-1$. For $m\leq r- k$ this does not happen
and the closest to $\T$ pole $z=t^{-1}w^{-1}$ is reached for $m=r-k$, $k>0$.
The $t^{-1}w$-point related poles from $P_{\rm out}^B$ cannot be cancelled and the closest to
$\T$ pole is $z=t^{-1}wq^{-1}$, reached for $m-k=r-1$.

As a result of this analysis we see that for $s\to 0$ only two pairs of poles
pinch the contour of integration. These are $u=s+v,-s+v$, reached for $m=k$,
and $u=s-v,-s-v$, reached for $m=\tilde k$, $\tilde k=r-k,\, k>0$, and $\tilde k=0$ for $k=0$.
Deforming the cut $[-\frac{1}{2},\frac{1}{2}]$ to the contour $C$ which
crosses two poles $u=s\pm v$ and applying the Cauchy theorem we obtain
\begin{eqnarray}\nonumber
&& M(s,0)_{v,k;u,m}f(u,m):=
\frac{1}{2}(p^r;p^r)_\infty(q^r;q^r)_\infty
\\  \nonumber && \makebox[2em]{} \times
\int_C\sum_{m\in\mathbb{Z}_r+\mu}
\frac{\Gamma(s\pm v \pm u, \pm k \pm m;p,q)}
{\Gamma(2s,0)\Gamma(\pm2u,\pm 2m;p,q)} f(u,m)du
\\ && \makebox[2em]{}
+\frac{1}{2}\left( \frac{\Gamma(-2v,-2k)}{\Gamma(-2s-2v,-2k)}f(s+v,k)
+ \frac{\Gamma(2v,2k)}{\Gamma(-2s+2v,2k)}f(s-v,\tilde k)\right).
\nonumber\end{eqnarray}
Actually, in the discrete arguments of gamma functions emerging in the last term there
are various shifts by multiples of $r$ caused by the substitution $m=r-k$. They do not play
role because of the periodicity property \eqref{period}. However, if we would be using
the quasiperiodic function  \eqref{regm} then such shifts would bring nontrivial
multipliers creating inconveniences.
Now we can take the limit $s\to 0$. The integral over the contour $C$ remains finite
because there are no singularities on the integration contour. However, the multiplier
$1/\Gamma(2s,0)$ vanishes in this limit and thus the integral term disappears.
The rest yields the desired statement \eqref{unit}.

We note that the integrand in \eqref{REFT} in general is not invariant under the shift
$u\to u+1$ due to nontrivial normalization factors for the elliptic gamma function.
Indeed, these factors yield the $u$-dependent multiplier of the form $e^{2\pi i u m(2n+3r)/r}$ ,
which clearly has brunch cuts in terms of the variable $z=e^{2\pi i u}$.
Only for special choices of the function $\alpha(u,m)$ the integrand is invariant under
the shift $u\to u+1$ (an example of such a choice related to the rarefied elliptic beta
integral will be given below). In the latter case the integrand becomes meromorphic in
$z\in \mathbb{C}^\times$, i.e. the branch cuts in $z$ disappear and one can pass to
contour integrals over $z$.
However, even in the absence of the periodicity, for any real analytic $\alpha(u,m)$
the integral over any finite interval of $u\in\mathbb{R}$ is well defined.
The cut $u\in [-\frac{1}{2},\frac{1}{2}]$ was chosen as the integration interval
in order to have the symmetry $u\to -u$ for the integrand without $\alpha(u,m)$-function.
Under the condition that $\alpha(u,m)=\alpha(-u,r-m),\, m\in \Z_r+\mu$, one
can reduce the summation in $m$ in \eqref{REFT} to the form
given in \eqref{e=0red} and \eqref{e=1red} for the rarefied elliptic beta integral.

We defined Bailey pairs in terms of the $r$-periodic $\Gamma$-function \eqref{REGper},
which is not single valued for variables $t, w, z$. One could use instead
 in the definition \eqref{REFT} the differently normalized single valued
 $\Gamma$-function \eqref{regm}, but as mentioned above, it would introduce
various normalization factors in the relations derived above and below because of the absence
of $r$-periodicity \eqref{period}. However,  both approaches  lead to equivalent star-triangle
relations and Yang-Baxter equations discussed below because of the restoration of
either singlevaluedness or $r$-periodicity of emerging integral kernels.

Consider now the rarefied analogue of the integral Bailey lemma.
Suppose that we have an explicit Bailey pair related by the equality \eqref{REFT}.
Then, the following transformations generate an infinite chain of Bailey pairs.
For $\alpha(v,k;t,n)$ we set
\beq
\alpha'(v,k;s+t,\ell+n)=D(s,\ell;y,j;v,k)\alpha(v,k;t,n),
\label{a'}\ee
where the coefficient
\beq
D(s,\ell;y,j;v,k):=\Gamma(\textstyle{\frac{\tau+\sigma}{2}}-s\pm y \pm v,-\ell\pm j\pm k),
\quad p(\ell)=p(j)p(k),
\label{D}\ee
obeys the properties
$$
D(s,\ell;y,j;v,k)D(-s,-\ell;y,j;v,k)=1,
$$
and $D(0,0;y,j;v,k)=1$ for $p(j)=p(k)$.
Because of the lack of alphabets, here we use the letter $t$ in additive meaning
(i.e. we replace in the multiplicative notation $t\to e^{2\pi i t}$).
For $\beta(v,k;t,n)$ we set
\begin{eqnarray} \nonumber 
&& \beta'(v,k;s+t,\ell+n)=D(-t,-n;y,j;v,k)
\\ && \makebox[4em]{}
\times  M(s,\ell)_{v,k;x,m}D(s+t,\ell+n;y,j;x,m)\beta(x,m;t,n).
\label{b'}\end{eqnarray}

From the demand $\beta'(v,k;s+t,\ell+n)=M(s+t,\ell+n)_{v,k;u,a}\alpha'(u,a;s+t,\ell+n)$,
we come to the operator identity
\begin{eqnarray} \nonumber &&
M(s,\ell)_{v,k;x,m}D(s+t,\ell+n;y,j;x,m)M(t,n)_{x,m;u,a}
\\ && \makebox[2em]{}
=D(t,n;y,j;v,k)M(s+t,\ell+n)_{v,k;u,a}D(s,\ell;y,j;u,a),
\label{RSTR}\end{eqnarray}
where for self-consistency one should choose $p(m)=p(a)p(k)p(j)$.
This is the braid ($MDM=DMD$) or operator star-triangle relation.
It is true due to the evaluation formula for the rarefied elliptic beta integral
\eqref{rfint}. Indeed, substituting explicit expressions for the integral
operator factors in \eqref{RSTR} one can check that the internal sum of integrals
over the variable $x$ on the left-hand side can be computed with the help
of the formula \eqref{rfint}. This yields exactly the right-hand side
integral operator expression. The change of the order of integration
over $x$ and $u$ is legitimate due to the Fubini theorem.

Let us apply the limiting relation \eqref{unit} to \eqref{RSTR} in the form $s\to -t$
taken for $n+\ell=0$. Then on the left-hand side
one has the factor $D(s+t,\ell+n;y,j;x,m)\to 1$ and on the right-hand side
$D(t,n;y,j;v,k)$ and $D(s,\ell;y,j;u,a)$ cancel each other due to the
property \eqref{unit}. Thus, for holomorphic functions $f(x,\ell)$
one gets the inversion relation
\beq
M(-t,-n)_{v,k;u,m}M(t,n)_{u,m;x,\ell}f(x,\ell)=\frac{1}{2}(f(v,k)+f(-v,\tilde k)).
\label{Minv}\ee
Here one has to specify the contours of integrations in the $M$-operators.
If $\text{Im} (t)>0$, then the contour in $M(-t,-n)_{v,k;u,m}$ should be deformed in
such a way that its sequences of poles converging to zero $P_{\rm in}^{A,B}$
are separated from the sequences going to infinity $P_{\rm out}^{A,B}$.
If $\text{Im} (t)<0$, then the contour in $M(t,n)_{u,m;x,\ell}$ should be deformed in
analogous way. Such contours exist if there is no pinching by the pairs of poles,
which may happen, e.g., if $2t=\tau j + \sigma k,\, j,k\in\Z^2.$
For $r=1$ the property \eqref{Minv} was established in \cite{SW}
for a particular restriction of parameters.

So, we have the situation resembling the Fourier transform when the change of signs
is equivalent to the inversion of the integral transform.

Let us denote in the rarefied elliptic beta integral evaluation formula \eqref{rfintadd}
the parameters $t_5=t+x,\, t_6=t-x,\, n_5=n+j,\, n_6=n-j$. This generates
the explicit Bailey pair of the form
\begin{align*} &
\alpha(u,m;t,n)=\prod_{a=1}^4\Gamma(t_a\pm u,n_a\pm m), \quad 2t+\sum_{a=1}^4t_a=\sigma+\tau,
\quad 2n+\sum_{a=1}^4n_a=0,
\\  &
\beta(x,j;t,n)=\prod_{a=1}^4\Gamma(t\pm x +t_a,n\pm j +n_a)\prod_{1\leq a<b\leq 4}\Gamma(t_a+t_b,n_a+n_b).
\end{align*}
Then the relation $\beta'(v,j;s+t,\ell+n)=M(s+t,\ell+n)_{v,j;u,m}\alpha'(u,m;s+t,\ell+n)$
yields the key $W(E_7)$-group symmetry transformation \cite{spi:rare}
\begin{eqnarray}\label{E7trafo1} &&  \makebox[-1em]{}
V^{(r)}(t_a,n_a)=
\prod_{1\leq b<c\leq 4}\Gamma (t_b+t_c,n_b+n_c)
\Gamma (t_{b+4}+t_{c+4},n_{b+4}+n_{c+4}) V^{(r)}(s_a,k_a),
\end{eqnarray}
for a rarefied elliptic analogue of the Euler-Gauss hypergeometric function
\beq
V^{(r)}(t_a,n_a)=\kappa^{(r)}
\sum_{m\in\Z_r+\mu}\int_{-1/2}^{1/2}
\frac {\prod_{a=1}^8\Gamma(t_a\pm u,n_a\pm m) }
{\Gamma (\pm 2u,\pm 2m)} du, \quad \text{Im}(t_a)>0, \; n_a\in\Z_r+\mu.
\label{V}\ee
The function $V^{(r)}(s_a,k_a)$ standing on the right-hand side of \eqref{E7trafo1}
has the arguments for $a=1,2,3,4$,
$$
s_a =t_a+t,\quad s_{a+4}=t_a-t, \quad t= \frac{1}{2}(\sigma+\tau- \sum_{b=1}^4t_b)
$$
and
$$
k_a= n_a+n,\quad k_{a+4}= n_a-n,\quad n=-\frac{1}{2}\sum_{b=1}^4n_b=\frac{1}{2}\sum_{b=5}^8n_b.
$$
It has the same form as \eqref{V},
but the summation goes now over $m\in\Z_r+\nu$, where $\nu=0, \frac{1}{2}$ is determined
from the condition $k_a\in\Z_r+\nu$ with
$$
p(k_a)=(-1)^{2\nu}=(-1)^{2(n_a+n)}=p(n_a)\prod_{b=1}^4(-1)^{n_b}.
$$
Thus, the summation in $m$ for  $V^{(r)}(s_a,k_a)$ goes over $\Z_r$ or $\Z_r+1/2$
depending on whether the sum $2\mu+\sum_{b=1}^4n_b$ is even or odd, as it has
been found in  \cite{spi:rare}.

\section{Connection to lattice spin systems}

Consider now a square lattice and place at its vertices the ``spins'' with the
continuous $x\in \C$ and discrete $m\in\Z_r+\mu$ components.
Denote  $pq=e^{-4\pi\eta}$, i.e. $\eta=-i(\tau+\sigma)/2$,
and $s=i\alpha$, $t=i\beta$ in relations \eqref{a'}-\eqref{Minv}.
Now we introduce the Boltzmann weight for horizontal edges connecting neighboring
spins of the lattice
\beq
W_{\alpha,\ell}(v,k;x,m)=\Gamma(i(\eta-\alpha)\pm v\pm x,-\ell\pm k \pm m)=D(i\alpha,\ell;v,k;x,m),
\label{Boltzmann}\ee
where we call $\alpha$ and $\ell$ the rapidities.
Clearly, it obeys the inversion property $W_{\alpha,\ell}W_{-\alpha,-\ell}=1$.
To the vertical edges we ascribe the Boltzmann weight $W_{\eta-\alpha,-\ell}(v,k;x,m)$.

Acting by the operator star-triangle relation onto the combination of Dirac delta functions
$\delta(u-z)\delta_{a-h,0}+\delta(u+z)\delta_{a-\tilde h,0}$ (where, $\tilde h=0$ for $h=0$
and $\tilde h=r-h$ for $h>0$).
Since the essential part of
the kernel of $M$-operator is described by the $W$-function, we obtain the equality
\begin{eqnarray}\nonumber &&
\int_0^{1}\sum_{m\in\Z_r+\mu}\rho(x,m)W_{\eta-\alpha,-\ell}(v,k;x,m)
W_{\alpha+\beta,\ell+n}(y,j;x,m)W_{\eta-\beta,-n}(z,h;x,m)dx
\\ && \makebox[2em]{}
=\chi(\alpha,\beta)W_{\beta,n}(y,j;v,k)
W_{\eta-\alpha-\beta,-\ell-n}(v,k;z,h)W_{\alpha,\ell}(y,j;z,h),
\label{STRfree}\end{eqnarray}
where
\beq
\chi(\alpha,\beta)=\Gamma(2i\alpha,2\ell)\Gamma(2i\beta,2n)\Gamma(2i(-\alpha-\beta+\eta),-2\ell-2n).
\label{chi}\ee
and
\begin{eqnarray}\nonumber &&
\rho(x,m)=\frac{(p^r;p^r)(q^r;q^r)}{2\Gamma(\pm 2x,\pm 2m;p,q)}
\\ && \makebox[2em]{}
=\frac{1}{2}(p^r;p^r)_\infty(q^r;q^r)_\infty (pq)^{\frac{m(2m-r)}{2r}}\theta(e^{4\pi i x}p^{2m};p^r)
\theta(e^{-4\pi i x}q^{2m};q^r).
\label{rho}\end{eqnarray}
For the taken spin lattice system the function $\rho(x,m)$ describes a self-energy of the spins
sitting at the vertices and \eqref{STRfree} is the standard functional form of the star-triangle
relation guaranteeing integrability of the model.

For $r=1$, when the discrete components of the spin
variables are absent, the described lattice spin system was considered in \cite{BS}. For
arbitrary $r$ and $\mu=0$ it was formulated in \cite{kels}. In general discrete spins form such
star-triangle relations were considered in \cite{bax:PLA} (see also \cite{bax:oda}).
In the latter case the spins were taking discrete values, self-energies of the spins were suppressed
and the normalization factor $\chi$ was denoted as $R_{pqr}$ ($p,q,r$ are the rapidities and in the
models described here $R_{pqr}$ has the factorized form of dependence  on the rapidities \cite{bax:oda}).

Let us renormalize the Boltzmann weight
$$
W_{\alpha,\ell}(y,j;u,h)=m(\alpha,\ell)\widetilde W_{\alpha,\ell}(y,j;u,h)
$$
and choose, if it is possible, the multiplier $m(\alpha,\ell)$ from the condition
\beq
m(\eta-\alpha,-\ell)=\Gamma(2i\alpha,2\ell)m(\alpha,\ell),\quad
m(-\alpha,-\ell)m(\alpha,\ell)=1.
\label{norm_eq}\ee
Then in terms of the $\widetilde W$-weight one gets the star-triangle relation without
$\chi$-multiplier on the right-hand side
\begin{eqnarray}\nonumber &&
\int_0^{1}\sum_{m\in\Z_r+\mu}\rho(x,m)\widetilde W_{\eta-\alpha,-\ell}(v,k;x,m)
\widetilde W_{\alpha+\beta,\ell+n}(y,j;x,m)\widetilde W_{\eta-\beta,-n}(z,h;x,m)dx
\\ && \makebox[2em]{}
=\widetilde W_{\beta,n}(y,j;v,k)
\widetilde W_{\eta-\alpha-\beta,-\ell-n}(v,k;z,h)\widetilde W_{\alpha,\ell}(y,j;z,h).
\label{STR}\end{eqnarray}

The partition function of this spin system has the form
$$
Z=\sum \int \prod_{(ij)}\widetilde W_\alpha(u_i,m_i;u_j,m_j)
\prod_{(kl)}\widetilde W_{\eta-\alpha}(u_k,m_k;u_l,m_l)\prod_{a}\rho(u_a,m_a)du_a,
$$
where the products $\prod_{(ij)}$ and $\prod_{(kl)}$ are taken over the horizontal
and vertical edges, respectively, while the product over vertices $\prod_{a}$ takes
into account self-energies of the spins. According to \cite{BS}, for real and positive
Boltzmann weights the quantity $m(\alpha,\ell)$ appearing in the
representation \eqref{STR} may be interpreted as the partition function, if it has appropriate
analytical properties required by the inversion method of solving statistical mechanics
models \cite{Baxter,Stroganov,Zamolodchikov}. Correspondingly, the appropriately
normalized Boltzmann weight $ \widetilde  W_{\alpha,\ell}$ leads to the model when the free energy
per edge vanishes in the thermodynamic limit, $\lim_{N,M\to \infty} \frac{1}{NM} \log Z=0$,
for the $N\times M$ sites spin lattice.

As observed in \cite{stat},
the partition function $Z$ coincides with the superconformal index
of a particular four-dimensional supersymmetric quiver gauge theory, whereas the
star-triangle relation and Yang-Baxter equation (the integrability conditions)
describe certain electromagnetic dualities of gauge theories with a simple
gauge group. This yields an interesting $4d/2d$ correspondence picture
 (or $3d/2d$, $2d/2d$ correspondences in the degenerate cases) between
 the gauge field theories and $2d$ statistical mechanics.

Physical values of spins in the Boltzmann weights $W_{\alpha,\ell}(y,j;v,k)$
are $y, v\in [0,1]$ and $j,k\in \Z_r+\mu$. For $r>1$ the weight $W_{\beta,n}(y,j;v,k)$
is real and positive if $p^*=q$, $0<\alpha<\eta$ and, moreover, only for vanishing
discrete rapidity variable, $n=0$.
In this case the solution of equation \eqref{norm_eq} with $\ell=0$ was found
in \cite{kels} (it was done for the $\mu=0$ case, but the answer does not
depend on $\mu$):
\beq
m(\alpha,0)=\frac{\Gamma(e^{-4\pi\alpha}pq;p^r,q^r,(pq)^2)}
{\Gamma(e^{4\pi\alpha}pq;p^r,q^r,(pq)^2)}
=\exp\Big(\sum_{m\in\Z/\{0\}}\frac{(pqe^{4\pi\alpha})^n(1-(pq)^{rn})}
{n(1-(pq)^{2n})(1-p^{rn})(1-q^{rn})}\Big),
\label{m}\ee
where $\Gamma(z;p,q,t)$ is the second order elliptic gamma function.
For $r=1$ this solution was obtained in \cite{BS}
(in this case the weight is real and positive also when $p$ and $q$ are real).
In a particular limit one finds similar normalization factors in
the star-triangle relations for the Faddeev-Volkov model \cite{BMS}
and its generalization described in \cite{stat}.

We would like to draw attention to the fact that general solution of the equations
$$
f(\eta-\alpha)=\Gamma(2i\alpha,0)f(\alpha),\quad f(-\alpha)f(\alpha)=1
$$
satisfied by the normalization factor $m(\alpha,0)$ \eqref{norm_eq} has the form
$$
f(\alpha)=m(\alpha,0) e^{\varphi(\alpha)},
$$
where $m(\alpha,0)$ is given above and the function ${\varphi(\alpha)}$ satisfies
the equations
$$
\varphi(\eta-\alpha)=\varphi(\alpha),\quad \varphi(-\alpha)+\varphi(\alpha)=0,
$$
i.e.
\beq
\varphi(\alpha)=\sum_{k=0}^\infty c_k\sin \frac{\pi(2k+1)\alpha}{\eta},
\label{phi}\ee
where $c_k$ are arbitrary coefficients guaranteeing convergence of the series.
For real $c_k, \eta$ one has real and positive $m(\alpha) e^{\varphi(\alpha)}$.
The functional freedom $\varphi(\alpha)$ must be removed to claim that $m(\alpha,0)$
determines indeed the true asymptotics of the partition function.
As V.V.~Bazhanov clarified to the author, this should be possible to do from fixing
the analytical properties of the partition function and its asymptotics for Im$(\alpha)\to \infty$
which would rule out \eqref{phi} and fix \eqref{m} as a correct solution.
Such a rigorous analysis was not performed yet for the spin lattice models discussed above.

\section{The Yang-Baxter equation}

Algebraic relations for operator entries of the Bailey lemma can be represented in the form
of Coxeter relations for a permutation group.
Let us take six pairs of continuous and discrete variables
$a_j=(t_j,n_j)$, $\, j=1,\ldots,6$, $t_j\in\C,\, n_j\in\Z_r+\nu_j$, $\nu_j=0,\frac{1}{2}$.
Denote as $s_k,\, k=1,\ldots,5$,
elementary transposition operators generating the permutation group $\mathfrak{S}_6$:
$$
s_j(\mathbf{a}):=s_j(\ldots,a_j,a_{j+1},\ldots)=(\ldots,a_{j+1},a_j,\ldots), \quad j=1,\ldots, 5.
$$
Similarly we introduce three spin variables $\sigma_k=(u_k, m_k)$, $k=1,2,3,$ with
$u_k\in [-\frac{1}{2},\frac{1}{2}]$, $m_k\in \Z_r+\mu_k$, $\mu_k=0,\frac{1}{2}$.

We introduce five operators $\mathrm{S}_{1}(\mathbf{a}),\ldots,\mathrm{S}_{5}(\mathbf{a})$
acting on the functions of three spins $f(\sigma_1,\sigma_2,\sigma_3)$ as follows
\begin{eqnarray} \nonumber &&
[\mathrm{S}_1(\mathbf{a})f](\sigma_1,\sigma_2,\sigma_3):=M(a_1-a_2)_{\sigma_1,\sigma}f(\sigma,\sigma_2,\sigma_3), \quad
\\    \nonumber&&
[\mathrm{S}_2(\mathbf{a})f](\sigma_1,\sigma_2,\sigma_3):=D(a_2-a_3;\sigma_1,\sigma_2)f(\sigma_1,\sigma_2,\sigma_3),
\\ &&
[\mathrm{S}_3(\mathbf{a})f](\sigma_1,\sigma_2,\sigma_3):=M(a_3-a_4)_{\sigma_2\sigma}f(\sigma_1,\sigma,\sigma_3),
\\ &&
[\mathrm{S}_4(\mathbf{a})f](\sigma_1,\sigma_2,\sigma_3):=D(a_4-a_5;\sigma_2,\sigma_3)f(\sigma_1,\sigma_2,\sigma_3),
 \nonumber  \\ && 
[\mathrm{S}_5(\mathbf{a})f](\sigma_1,\sigma_2,\sigma_3):=M(a_5-a_6)_{\sigma_3,\sigma}f(\sigma_1,\sigma_2,\sigma),
 \nonumber\end{eqnarray}
where parities of discrete variables are correlated as $p(n_1)p(n_2)=p(m_1)p(m)$, etc.
Here we use the short-hand notation $a_j-a_k=(t_j-t_k,n_j-n_k)$,
$M(a)_{\sigma_j,\sigma}f(\sigma)=M(t,n)_{x_j,n_j;x,m}f(x,m)$
and $D(a;\sigma_1,\sigma_2)=D(t,n;u_1,m_1; u_2,m_2)$.
Products of these operators take into account twisting of the operator arguments
\beq
\mathrm{S}_j\mathrm{S}_k:=\mathrm{S}_j(s_k \mathbf{a})\mathrm{S}_k(\mathbf{a}).
\label{mult}\ee
The function $f(\sigma_1,\sigma_2,\sigma_3)$ can be considered as a tensor product of
three identical infinite dimensional spaces $\V_1\otimes\V_2\otimes \V_3$ composed of sequences
of functions $f(u,m)\in\V$, $u\in\C$, $m\in\Z_r+\mu$.

Let us assume that the operators $\mathrm{S}_j$ are applied to sequences of
functions satisfying the constraint  $f(u,m)=f(-u,r-m),\, m\in \Z_r+\mu$,
for all spin variables $\sigma=(u,m)$.
Now it is not difficult to verify validity of the relations
\begin{equation}
\mathrm{S}_j^2=1, \quad \mathrm{S}_i\mathrm{S}_j=\mathrm{S}_j\mathrm{S}_i \
\text{ for } \ |i-j|>1, \quad
\mathrm{S}_j\mathrm{S}_{j+1}\mathrm{S}_j
=\mathrm{S}_{j+1}\mathrm{S}_j\mathrm{S}_{j+1}
\label{coxeter}\end{equation}
as a consequence of the identities following from the Bailey lemma.
Indeed, quadratic relations are equivalent to the inversion relations for
the $M$- and $D$-operators
$$
[\mathrm{S}_1^2f](\sigma_1)=[\mathrm{S}_1(s_1\mathbf{a})\mathrm{S}_1(\mathbf{a})f](\sigma_1)
=[M(a_2-a_1)M(a_1-a_2)f](u_1,m_1)=f(u_1,m_1),
$$
and
$$
\mathrm{S}_2^2=\mathrm{S}_2(s_2\mathbf{a})\mathrm{S}_2(\mathbf{a})=D(a_3-a_2)D(a_2-a_3)=1,
$$
where we  omit spin variable labels for $M$- and $D$-operators.

The cubic relations are equivalent to the star-triangle relation \eqref{STR} based on
the rarefied elliptic beta integral \eqref{rfint}, e.g.
\begin{eqnarray} \nonumber
&&
\mathrm{S}_1\mathrm{S}_2\mathrm{S}_1
=\mathrm{S}_1(s_2s_1\mathbf{a})\mathrm{S}_2(s_1\mathbf{a})\mathrm{S}_1(\mathbf{a})
=M(a_2-a_3)D(a_1-a_3)M(a_1-a_2)\\
&&
=D(a_1-a_2)M(a_1-a_3)D(a_2-a_3)=
\mathrm{S}_2(s_1s_2\mathbf{a})\mathrm{S}_1(s_2\mathbf{a})\mathrm{S}_2(\mathbf{a})
=\mathrm{S}_2\mathrm{S}_1\mathrm{S}_2.
\nonumber\end{eqnarray}

Let us take operators $\mathrm{S}_i$ introduced above and define
two composite operators called R-matrices. The first one $\mathrm{R}_{12}(\mathbf{a})$
acts nontrivially in the space $\V_1\otimes\V_2$ and has the form
\begin{eqnarray}\nonumber  &&
\mathrm{R}_{12}(\mathbf{a}) =\mathrm{R}_{12}(a_1,a_2|a_3,a_4)
=\mathrm{S}_2(s_1s_3s_2\mathbf{a})
\mathrm{S}_1(s_3s_2\mathbf{a})\mathrm{S}_3(s_2\mathbf{a})
\mathrm{S}_2(\mathbf{a})
\\  \makebox[2em]{} &&
= \mathrm{S}_2(a_1-a_4)
\mathrm{S}_1(a_1-a_3)\mathrm{S}_3(a_2-a_4)\mathrm{S}_2(a_2-a_3).
\label{R} \end{eqnarray}
This R-matrix corresponds to an operator form representation of the checkerboard (IRF)
lattice Boltzmann weights \cite{bax:PLA}. Such factorizations of the general R-matrices
for models with the continuous spin variables were discussed earlier in \cite{SD,DM}.

The second R-matrix $\mathrm{R}_{23}(\mathbf{a})$
acts nontrivially in the space $\V_2\otimes\V_3$
\begin{eqnarray}\nonumber  &&
\mathrm{R}_{23}(\mathbf{a})=\mathrm{R}_{23}(a_3,a_4|a_5,a_6) =
\mathrm{S}_4(s_3s_5s_4\mathbf{a})
\mathrm{S}_3(s_5s_4\mathbf{a})\mathrm{S}_5(s_4\mathbf{a})
\mathrm{S}_4(\mathbf{a})
\\ && \makebox[4em]{}
=\mathrm{S}_4(a_3-a_6)\mathrm{S}_3(a_3-a_5)\mathrm{S}_5(a_4-a_6)
\mathrm{S}_4(a_4-a_5).
 \label{R23}\end{eqnarray}
These operators are called R-matrices and they satisfy the following Yang-Baxter relation
\begin{eqnarray} \nonumber &&
\mathrm{R}_{23}(a_{1},a_{2}|a_3,a_4)\,\mathrm{R}_{12}(a_1,a_{2}|a_5,a_6)\,
\mathrm{R}_{23}(a_3,a_4|a_5,a_6)\,
\\ &&
=
\mathrm{R}_{12}(a_3,a_4|a_5,a_6)\,\mathrm{R}_{23}(a_{1},a_2|a_5,a_6)\,
\mathrm{R}_{12}(a_1,a_2|a_3,a_4).
\label{RRR}\end{eqnarray}
After replacing R-matrices by their factorized expressions in terms of $\mathrm{S}_k$-operators
this equation can be written in the following form
\beq
\mathrm{S}_4\mathrm{S}_3\mathrm{S}_2\mathrm{S}_3\cdot
\mathrm{S}_1\mathrm{S}_5\mathrm{S}_4\mathrm{S}_5\cdot
\mathrm{S}_3\mathrm{S}_2\mathrm{S}_3\mathrm{S}_4
=
\mathrm{S}_4\mathrm{S}_3\mathrm{S}_5\mathrm{S}_4\cdot
\mathrm{S}_2\mathrm{S}_1\mathrm{S}_3\mathrm{S}_2\cdot
\mathrm{S}_4\mathrm{S}_3\mathrm{S}_5\mathrm{S}_4.
\label{YBEproof}\ee
The proof of this identity follows from the cubic Coxeter relations \eqref{coxeter},
which imply the multiplication rule \eqref{mult},
and copies similar proofs presented in \cite{DM,DS}. We stress that in this proof
one uses only the braid group relations described by
the star-triangle relation \eqref{STR} and the commutativity  of
$\mathrm{S}_i$ and $\mathrm{S}_j$ for $|i-j|>1$.
So, in this picture the Yang-Baxter equation gets an interpretation of a word
identity in the group algebra of the permutation group $\mathfrak{S}_6$.

Let us define the operators $\PP_{ij}$ permuting spaces of spin variables
$$
\PP_{ij}f(\ldots,\sigma_i,\ldots,\sigma_j,\ldots)=f(\ldots,\sigma_j,\ldots,\sigma_i,\ldots).
$$
With their help we can write
$$
\mathrm{R}_{23}(\mathbf{a})= \PP_{13}\PP_{12}\mathrm{R}_{12}(a_3,a_4|a_5,a_6)\PP_{12}\PP_{13}.
$$
Let us define one more R-matrix
$$
\mathrm{R}_{13}({\bf a})= \mathrm{R}_{13}(a_1,a_2|a_5,a_6)
= \PP_{12}\mathrm{R}_{23}(a_1,a_2|a_5,a_6)\PP_{12}
$$
acting nontrivially in the product $\V_1\otimes\V_3$.

Parametrization of the R-matrices $R_{ij}$ by $a_1,\ldots, a_6$ is somewhat unusual.
The standard variables used for this purpose are the spectral variables
$u, v, w$ and the representation parameters $g_1, g_2, g_3$ (also called spins in the spin chains
formalism) related to $a_j$ as follows
\begin{eqnarray*} &&
a_1=\frac{u+ g_1}{2}, \quad a_{2}=\frac{u- g_1}{2}, \quad a_{3}=\frac{v+ g_2}{2},
\quad  a_4=\frac{v-g_2}{2},
\\  &&  \makebox[7em]{}
a_5=\frac{w+ g_3}{2}, \quad a_6=\frac{w-g_3}{2}.
\end{eqnarray*}
Here all variables consist of two components --- the continuous and discrete ones,
i.e. $u=(\lambda_u,k_u),\, v=(\lambda_v,k_v),\, w=(\lambda_w,k_w),$ and $g_j=(c_j,l_j)$.

Multiply now the left-hand side expression in \eqref{RRR} by the operator
$\PP_{12}\PP_{13}\PP_{23}$ and the right-hand side expression by the equal operator
$\PP_{23}\PP_{13}\PP_{12}$.
Pulling permutation operators $\mathbb{P}_{jk}$ to appropriate R-matrices
we come to the Yang-Baxter equation of the standard form
\begin{equation}\label{YB}
\mathbb{R}_{12} (u-v)\,\mathbb{R}_{13}(u-w)\, \mathbb{R}_{23}(v-w)
=\mathbb{R}_{23}(v-w)\,\mathbb{R}_{13}(u-w)\,\mathbb{R}_{12}(u-v),
\end{equation}
where we denoted
\beq
\mathbb{R}_{12} (u-v)=\mathbb{R}_{12} (u-v;g_1,g_2)= \mathbb{P}_{12}\mathrm{R}_{12}(\mathbf{a}),
\label{YBE_standard}\ee
and similarly
$$
\mathbb{R}_{23} (v-w)=\mathbb{P}_{23}\mathrm{R}_{23}(\mathbf{a}), \quad
\mathbb{R}_{13} (u-w)=\mathbb{P}_{13}\mathrm{R}_{13}(\mathbf{a}).
$$
We remind that each R-matrix $\mathbb{R}_{jk}$ in \eqref{YB} acts in the space $\V_j\otimes \V_k$,
where with each space $\V_j$ one associates the spectral and spin variables $u_j$ (i.e. $u, v, w$)
and $g_j$ encoded into the parameters $a_{2j-1}, a_{2j}$. Note, however, that after stripping off the
permutation variables $\PP_{12}$ the Yang-Baxter equation takes the form \eqref{RRR} where
the R-matrices $\mathrm{R}_{jk}$, acting in the same space
$\V_j\otimes \V_k$, depend on a different sets of spectral and spin variables.
For simplicity in \eqref{YB} we omitted dependence of $\mathbb{R}$-matrices on the spin variables $g_j$.
In the context of spin chains the spin variables $g_j$ are chosen to be equal for
all spaces $\V_j$ which justifies the latter convention.

Using the properties of $\mathrm{S}_j$-operators one can verify the
unitarity relations for these R-matrices
\begin{align}
\mathbb{R}_{ij}(u;g_i,g_j)\,\mathbb{R}_{ij}(-u|g_j,g_i) = 1.
\end{align}
In comparison to the standard Yang-Baxter equation,
R-matrices in the relation \eqref{YBE_standard} depend not only
on the difference of continuous spectral variables,
but also on a similar difference of discrete spectral variables,
$u-v=(\lambda_u-\lambda_v, k_u-k_v)$.

The derived R-matrix has the following explicit form
\begin{eqnarray} \nonumber &&
[\mathbb{R}_{12}(a_1,a_2|a_3,a_4)f](\sigma_1,\sigma_2)=(\kappa^{(r)})^2
\Gamma\big(\textstyle{\frac{\sigma+\tau}{2}}+t_1-t_4\pm x_1\pm x_2,n_1-n_4\pm m_1\pm m_2\big)
\\ \nonumber && \makebox[2em]{} \times
\sum_{k_j\in \Z_r+\mu_j}\int_{-1/2}^{1/2}
\frac{\Gamma(t_3-t_1 \pm x_2\pm u_1, n_3-n_1\pm m_2\pm k_1)}
{\Gamma(2(t_3-t_1),2(n_3-n_1))}
\\ \nonumber && \makebox[8em]{} \times
\frac{\Gamma(t_4-t_2\pm x_1\pm u_2, n_4-n_2\pm m_2\pm k_2)}
{\Gamma(2(t_4-t_1), 2(n_4-n_1))}
\\ && \makebox[2em]{} \times
\frac{\Gamma(\frac{\sigma+\tau}{2}+t_2-t_3\pm u_1\pm u_2,n_2-n_3\pm k_1\pm k_2)}
{\Gamma(\pm 2u_1,\pm 2k_1)\Gamma(\pm2 u_2, \pm 2 k_2)}f(u_1,k_1;u_2,k_2)du_1du_2.
\label{Rexplicit}\end{eqnarray}
For $r=1$ this R-matrix reduces to the one derived in \cite{DS}.
Since the whole formalism for $r>1$ is quite similar to the $r=1$ one,
it is natural to pose the task of building finite-dimensional
solutions of the Yang-Baxter equation reducing for $r=1$ to those
obtained in \cite{CDS}.

\smallskip

This work is supported by the Russian Science Foundation
(project no. 19-11-00131).
The author is indebted to V. V. Bazhanov and S. E. Derkachov for useful discussions
and to the referees for helpful remarks.

\end{document}